# The new science of complexity

Joseph L. McCauley
Physics Department
University of Houston
Houston, Texas 77204

jmccauley@uh.edu

key words:
complexity
simulations
laws of nature
invariance principles
socio-economic sciences

**Abstract**

Deterministic chaos, and even maximum computational complexity, have been discovered within Newtonian dynamics. Encouraged by comparisons of the economy with the weather, a Newtonian system, economists assume that prices and price changes can also obey abstract mathematical laws of motion. Meanwhile, sociologists and other postmodernists advertise that physics and chemistry have outgrown their former limitations, that chaos and complexity provide new holistic paradigms for science, and that the boundaries between the hard and the soft sciences, once impenetrable, have disappeared along with the Berlin Wall. Three hundred years after the deaths of Galileo, Descartes, and Kepler, and the birth of Newton, reductionism would appear to be on the decline, with holistic approaches to science on the upswing. We therefore examine the evidence that dynamical laws of motion may be discovered from empirical studies of chaotic or complex phenomena, and also review the foundations of reductionism.

**Socio-economic fields and "system theory"**

I define "system theory" to include mathematical models written in terms of systems of deterministic and stochastic ordinary and partial differential equations, iterated maps, and deterministic and stochastic automata. The idea is to include every possible kind of dynamical modelling.

In attempts to describe socio-economic phenomena from the standpoint of system theory it is Platonically assumed that the probability distributions describing prices and price changes, or other social factors, are determined by an objective mathematical law that governs how the economic system evolves [1]. This assumption is not only sufficient but is also necessary if the idea of mathematical law in economics is to make any sense. In physics and chemistry the ideas of entropy, thermodynamics, and nonequilibrium statistical mechanics are grounded in universally-valid microscopic dynamics. Without the underlying dynamics of particles, fluids and solid or plastic bodies there would be no dynamical origin for macroscopic probability distributions.

By a mathematical law of nature I mean a law of motion, a mathematical law of time-evolution. Galileo and Kepler discovered the simplest special cases. Their local laws were generalized by Newton to become three universally-valid laws of motion, along with a universal law of gravity. Newton's laws are "universal" in the following sense: they can be verified, often with very high decimal precision, regardless of where and when on earth (or on the moon or in an artificial satellite) careful, controlled experiments, or careful observations, are performed. It is the main purpose of this paper to stress the implications of the fact that no comparable result has ever been found in the socio-economic fields.

"Laws" of economics, "laws" of human behavior, and the Darwin-Wallace "laws" of fitness, competition, selection and adaptation are sometimes mentioned in the same context as laws of motion of inanimate matter (physics and chemistry), although since the time of Galileo the word "law" in the first three cases does not have the same import as in the case of physico-chemical phenomena. Confusion over what constitutes a law of nature is ancient: Aristotle invented a purely qualitative, holistic approach to the description of nature. Not recognizing any distinction between the different uses of the idea of natural law, he lumped together as "motion" the rolling of a ball, the education of a boy, and the growth of an acorn [2]. Ibn-Rushd realized that Aristotle's philosophy is consistent with a purely mechanistic picture of the universe. The growing influence of the mechanistic interpretation of nature in western Europe set Tomasso d'Aquino into motion in the thirteenth century. Aristotle did not use mathematics, but mechanism and mathematics go hand in hand. Is *human* nature, in some still-unknown mathematical sense, also mechanistic?

In the first chapter of his text on elementary economics [3], Samuelson tries to convince both the reader and himself that the difference between the socio-economic fields and the laws of physics is blurry, so that economics can be treated as if it would also be a science subject to mathematical law. Samuelson claims that physics is not necessarily as lawful as it appears, that the laws of physics depend subjectively on one's point of view. His argument is based on a nonscientific example of ambiguity from the visual perception of art (figure 1), and is genetically related through academic mutation and evolution to a viewpoint that has been advanced by the postmodernist and deconstructionist movement in art, literature, philosophy, psychology, and sociology. The latter argue that a text has no more meaning than the symbols on a printed page, that there is no universal truth, and therefore no universal laws of nature, and that Platonic-Ptolemeic astronomy and Aristotelian physics are still just as valid as fields of scientific study as are physics and astronomy since Galileo and Kepler (who revived the spirit of Archimedes).

Samuelson notes that physics relies on controlled experiments, and adds that in the socio-economic fields it is generally impossible to perform controlled experiments. This is not an excuse for bad science: controlled experiments are also impossible in astronomy where mathematical laws of nature have been verified with high decimal precision. See also Feynman [4] for criticism of the lack of isolation of cause and effect in the psycho-social fields.

Platonists in mathematics [5] form another category, believing that mathematical laws exist objectively and govern everything that happens. Physics is neither Aristotelian (qualitative and "holistic") nor Platonic (relying upon wishful thinking, because the "expected" mathematical laws are not grounded in careful, repeatable empiricism).

The divorce of the study of nature from Platonic and Aristotelian notions was initiated by Galileo and Descartes [6], but that divorce was not complete: with Galileo's empirical discoveries of two local laws of nature, the law of inertia and the local law of gravity, physics became a precise mathematico-empirical science. Biology, excepting the study of heredity since Mendel and excepting biochemistry and biophysics since the advent of quantum mechanics, has continued through the age of Darwin and beyond as a largely descriptive science in the tradition of Aristotle, with reliance upon vague, mathematically-undefined notions like "competition, natural selection and adaptation".

I will explain why economic and other social phenomena lie beyond the bounds of understanding from the standpoint of dynamical modelling that attempts to describe the time-evolution of systems, *even if the goal is merely to extract the crudest features like coarsegrained statistics.* I will give reasons why mathematical laws of economics do not exist in any empirical or computationally-effective [7] sense. In order to make my argument precise, I first review some little-known and poorly-understood facts about deterministic dynamical systems that include Newton's laws of motion for particles and rigid bodies, and also nondiffusive chemically-reacting systems.

**What does "nonintegrable" mean?**

We expect that any system of ordinary differential equations generating critical (orbitally-metastable), chaotic (orbitally-unstable), or complex dynamics must be both nonlinear and nonintegrable. Most of us think that we can agree on the meaning of "nonlinear". Before asking "What is complexity?" we first define what "nonintegrable" means [8,9].

The ambiguity inherent in both serious and superficial attempts to distinguish "integrability" from "nonintegrability" was expressed poetically by Poincaré, who stated that a dynamical system is generally neither integrable nor nonintegrable, but is more or less integrable [10]. For most scientists the explanation of various roots to chaos (via period doubling, e.g.) has tended to submerge rather than clarify the question how to distinguish those two ideas, but without eliminating many misconceptions. Modern mathematicians have managed to give some precise definitions of nonintegrability [11] that are hard to translate into simpler mathematics. Here, I try to describe what "nonintegrability" means geometrically and analytically.

For the sake of precision I frame my discussion in the context of flows in phase space,

$$\frac{dx}{dt} = V(x), \qquad (1)$$

where phase space is a flat inner product space so that the n axes labeled by $(x_1,...,x_n)$ can be regarded as Cartesian [12], and $V(x)$ is an n-component time-independent velocity field. Newtonian dynamical systems can always be rewritten in this form whether or not the variables $x_i$ defining the system in physical three dimensional space are Cartesian (for example, it is allowed have $x_1 = \theta$ and $x_2 = d\theta/dt$, where $\theta$ is an angular variable). Flows that preserve the Cartesian volume element $d\Omega = dx_1...dx_n$ are defined by $\nabla \cdot V = 0$ (conservative flows) while driven dissipative-flows correspond to $\nabla \cdot V \neq 0$, where $\nabla$ denotes the Cartesian gradient in n dimensions.

For a velocity field whose components satisfy the condition $V_1 + ... + V_n = 0$, then the global conservation law $x_1 + ... + x_n = C$ follows. This abstract case includes chemically-reacting systems with concentration $x_i$ for species i.

For a flow and for any initial condition $x_o$ the solution $x_i(t) = U(t)x_{io}$ has no *finite* time singularities [13] because singularities of trajectories of flows are confined to the complex time plane: the time evolution operator $U(t)$ exists for all real finite times t and defines a one-parameter transformation group with inverse $U^{-1}(t) = U(-t)$, so that one can in principle integrate backward in time, $x_{oi} = U(-t)x_i(t)$, as well as forward. In other words, even driven-dissipative flows are perfectly time-reversible.

Many researchers use floating point arithmetic in numerical integrations of chaotic systems but uncontrollable errors are introduced into numerical integrations by the use of floating point arithmetic, and those errors violate time reversibility in the simplest of cases. Even for a *nonchaotic* driven-dissipative flow floating-point errors will prevent accurate numerical solutions either forward or backward in time after only a relatively short time [12]. The simplest example is given by the one dimensional flow $dy/dt = y$, all of whose streamlines have the positive Liapunov exponent $\lambda = 1$ forward in time, and the negative Liapunov exponent $\lambda = -1$ backward in time. Consequently, the simple linear equation $dy/dt = y$ cannot be integrated forward in time accurately numerically, for moderately-long times, if floating point arithmetic is used.

Chaotic unimodal maps $z_n = f(z_{n-1})$ like the logistic map $f(x) = Dx(1-x)$ have a multi-valued inverse $z_{n-1} = f^{-1}(z_n)$ and therefore are not uniquely time-reversible. Contrary to superficial appearances based upon an unwarranted extrapolation of a numerical calculation, time reversal is not violated by the Lorenz model

$$\frac{dx_1}{dt} = \sigma(x_2 \pm x_1)$$

$$\frac{dx_2}{dt} = \rho x_1 \pm x_2 \pm x_1 x_3$$

$$\frac{dx_3}{dt} = \pm \beta x_3 + x_1 x_2 \qquad (1b)$$

in the chaotic regime. The well-known numerically-suggested one dimensional cusp map (figure 2) $z_n = f(z_{n-1})$ that represents maxima of a time series [14] of $x_3(t)$ at discrete times $t_0, t_1, ..., t_n, ...$ , where $t_n - t_{n-1}$ denotes the time lag between successive maxima $z_{n-1} = x_3(t_{n-1})$ and $z_n = x_3(t_n)$, can not have a double-valued inverse $z_{n-1} = f^{-1}(z_n)$: backward integration $z_{n-1} = U(t_{n-1} - t_n) z_n$ is unique for a flow, and the Lorenz model satisfies the boundedness condition for a flow [14]. Therefore, Lorenz's one dimensional cusp map $z_n = f(z_{n-1})$ is not continuous and may even be infinitely fragmented and nondifferentiable in order that the inverse map $f^{-1}$ doesn't have two branches. Note that the Lorenz model may describe a chemically-reacting system if $\beta = 0$ and $\sigma = \rho = 1$, in which case the flow is driven-dissipative but is not chaotic (the flow is orbitally-stable, with no positive Liapunov exponent in forward integration).

Surprise has been expressed that it was found possible to describe a certain chaotic flow by a formula in the form of an infinite series [8], but "nonintegrable" does not mean not solvable: any flow, even a critical, chaotic or complex one, has a unique, well-defined solution if the velocity field V(x) satisfies a Lipshitz condition (a Lipshitz condition requires the definition of a metric in phase space), or is at least once continuously differentiable, with respect to the n variables $x_i$. If, in addition, the velocity field is analytic in those variables then the power series

$$x_i(t) = x_{io} + t(Lx_i)_o + t^2(L^2x_i)_o/2 + ...., \quad (2)$$

where $L = V \cdot \nabla$, has a nonvanishing radius of convergence, so that the solution of (1) can in principle be described by the power series (2) combined with analytic continuation for *all* finite times [15]. It is well known that this is not a practical prescription for the calculation of trajectories at long times. The point is that a large category of deterministic chaotic and complex flows are precisely determined over any desired number of finite time intervals by *analytic formulae*. The Lorenz model (1b) provides an example. Analyticity is impossible for the case of truly "random" motion (like α-particle decays), where the specification of an initial condition does not determine a trajectory at all (as in Feynman's path integral), or for Langevin descriptions of diffusive motion, where almost all trajectories are also continuous and almost everywhere nondifferentiable (as in Wiener's functional integral).

According to Jacobi and Lie, a completely integrable dynamical system has n-1 global time-independent first integrals (conservation laws) $G_i(x_1,...,x_n) = C_i$ satisfying the linear partial differential equation

$$\frac{dG_i}{dt} = V \cdot \nabla G_i = V_k \frac{\partial G_i}{\partial x_k} = 0 \quad (3)$$

along any streamline of the flow. In addition, these conservation laws must (in principle, but not necessarily via explicit construction) determine n-1 "isolating integrals" of the form $x_k = g_k(x_n, C_1,...,C_{n-1})$ for k = 1,...,n-1. When all of this holds then the global flow is a time-translation for *all* finite times t in the Lie coordinate system

$$y_i = G_i(x_1,...,x_n) = C_i, \quad i = 1,...,n-1$$
$$y_n = F(x_1,...,x_n) = t + D \qquad (4)$$

defined by the n-1 conservation laws, and the system is called completely integrable. The solution reduces *in principle* to n *independent* integrations, and the flow is confined to a two-dimensional manifold that may be either flat or curved and is determined by the intersection of the n-1 global conservation laws. For the special case of a canonical Hamiltonian flow with f degrees of freedom, f commuting conservation laws confine the flow to a constant speed translation on an f dimensional flat manifold. The $n^{th}$ transformation function $F(x_1,...,x_n)$ is defined by integrating $dt = dx_n/V_n(x_1,...,x_n) = dx/v_n(x_n,C_1,...,C_{n-1})$ to yield $t + D = f(x_n,C_1,...,C_{n-1})$. One then uses the n-1 conservation laws to eliminate the constants $C_i$ in favor of the n-1 variables $x_i$ in f to obtain the function F. Whether one can carry out all or any of this constructively, in practice, is geometrically irrelevant: in the description (4) of the flow all effects of interactions have been eliminated globally via a coordinate transformation. The transformation (4) "parallelizes" (or "rectifies" [13]) the flow: the streamlines of (1) in the y-coordinate system are parallel to a single axis $y_n$ for all times, and the time evolution operator is a uniform time-translation $U(t) = e^{td/dy_n}$. Eisenhart asserted formally, without proof, that all systems of differential equations (1) are described by a single time translation operator [16], but this is possible globally (meaning for all finite times) only in the completely integrable case.

Although time-dependent first integrals are stressed in discussions of integrable cases of driven-dissipative flows like the Lorenz model [8], there is generally no essential difference between (3) and the case of n time-dependent first integrals $G'_i(x_1,...,x_n,t) = C'_i$ satisfying

$$\frac{dG_i}{dt} = V \cdot \nabla G_i + \frac{\partial G_i}{\partial t} = 0. \qquad (3b)$$

Relying on the implicit function theorem, one conservation law $G'_n(x_1,...,x_n,t) = C'_n$ can be used to determine a function $t = F'(x_1,...,x_n,C'_n)$, whose substitution into the other n-1 time-dependent conservation laws yields n-1 time-independent ones satisfying (3).

The n initial conditions $x_{io} = U(-t)x_i(t)$ of (1) satisfy (3b) and therefore qualify as time-dependent conservation laws, but initial conditions of (1) are generally only *trivial local* time-dependent conservation laws: dynamically seen, there is no qualitative difference between backward and forward integration in time. *Nontrivial* global conservation laws are provided by the initial conditions $y_{io}$, for i = 1, 2, ... , n-1, of a completely integrable flow in the Lie coordinate system (4), where the streamlines are parallel for all finite times: $dy_i/dt = 0$, i = 1,...,n-1, and $dy_n/dt = 1$.

Algebraic or at least analytic conservation laws [8] have generally been assumed to be necessary in order to obtain complete integrability. For example, Euler's description of a torque-free rigid body [12]

$$\frac{dL_1}{dt} = a L_2 L_3$$

$$\frac{dL_2}{dt} = \pm b L_1 L_3$$

$$\frac{dL_3}{dt} = c L_1 L_2 \quad , \quad (5)$$

with positive constants a , b , and c satisfying a - b + c = 0, defines a phase flow in three dimensions that is confined to a two dimensional sphere that follows from angular momentum conservation $L_1^2 + L_2^2 + L_3^2 = L^2$. Here, we have completely integrable motion that technically violates the naive expectation that each term in (4) should be given by a *single* function: for each period τ of the motion, the transformation function F has four distinct branches due to the turning points of the three Cartesian components $L_i$ of angular momenta on the sphere. In general, any "isolating integral" $g_k$ describing bounded motion must be multivalued at a turning point. Note also that the Lorenz model defines a certain linearly damped, driven symmetric top: to see this, set a = 0 and b = c = 1 in (5), and ignore all linear terms in (1b).

The few mathematicians who have discussed conservation laws in the literature usually have assumed that first integrals must be analytic or at least continuous [13] (however, see also ref. [11] where nonanalytic functions as first integrals are also mentioned). This is an arbitrary restriction that is not always necessary in order to generate the transformation (4) over all finite times: a two-dimensional flow in phase space, including a driven-dissipative flow, is generally integrable via a conservation law but that conservation law is typically singular. The conservation law is simply the function $G(x_1,x_2) = C$ that describes the two-dimensional phase portrait, and is singular at sources and sinks like attractors and repellers (equilibria and limit cycles provide examples of attractors and repellers in driven-dissipative planar flows) [12]. For the damped simple harmonic oscillator, for example, the conservation law has been constructed analytically [17] and is logarithmically singular at the sink. The planar flow where $dr/dt = r$ and $d\theta/dt = 0$ in cylindrical coordinates $(r,\theta)$ describes radial flow out of a source at $r = 0$. The conservation law is simply $\theta$, which is constant along every streamline and is undefined at $r = 0$. This integrable flow is parallelizeable for all *finite* times t simply by excluding one point, the source at $r = 0$ (infinite time would be required to leave or reach an equilibrium point, but the infinite time limit is completely unphysical). "Nonintegrable" flows do not occur in the phase plane. What can we say about "nonintegrability" about in three or more dimensions?

In differential equations [13] and differential geometry [18] there is also an idea of *local* integrability: one can parallelize an arbitrary vector field V about any "noncritical point", meaning about any point $x_o$ where the field V(x) does not vanish. The size $\varepsilon(x_o)$ of the region where this parallelization holds is *finite* and depends nonuniversally on the n gradients of the vector field. This means that we can "rectify" even chaotic and complex flows over a finite time, starting from any nonequilibrium point $x_o$. By analytic continuation [11,19], this local parallelization of the flow yields n-1 nontrival "local" conservation laws $y_i = G_i(x) = C_i$ that hold out to the first singularity of any one of the n-1 functions $G_i$, in agreement with the demands of the theory of first order linear partial differential equations (the linear partial differential equation (3) always has n-1 functionally independent solutions, but the solutions may be singular [17]).

Contemplate the trajectory of a "nonintegrable" flow that passes through any nonequilibrium point $x_o$, and let $t = 0$ when $x = x_o$. Let $t(x_o)$ then denote the time required for the trajectory to reach the first singularity of one of the conservation laws $G_k$. Such a singularity must exist, otherwise the flow would be confined for *all* finite times ("globally") to a single, smooth two-dimensional manifold. The global existence of a two-dimensional manifold can be prevented, for example, by singularities that make the n-1 conservation laws $G_i$ multivalued in an extension of phase space to complex variables [11]. Generally, as with solutions of (1) defined locally by the series expansion (2), the n-1 local conservation laws $G_i$ will be defined locally by infinite series with radii of convergence determined by singularities that lie in the complex extension of phase space. The formulae (4) then hold for a finite time $0 \leq t < t(x_o)$ that is determined by the distance from $x_o$ to the nearest complex singularity. Let $x_1(x_o)$ denote the point in phase space where that singularity causes the series defining $G_i$ to diverge. Following Arnol'd's [13] statement of the "basic theorem of ordinary differential equations", we observe that the streamline of a flow (1) passing through $x_o$ can not be affected by the singularity at $x_1(x_o)$ in the following superficial sense (consistent with the fact that the singularities of the functions $G_i$ are either branch cuts or phase singularities): we can *again* parallelize the flow about the singular point $x_1(x_o)$ and can *again* describe the streamline for *another* finite time $t(x_o) \leq t < t(x_1)$ by another set of parallelized flow equations of the form (4), where $t(x_1)$ is the time required to reach the next singularity $x_2(x_o)$ of any one of the n-1 conservation laws $G_i$, starting from the second initial condition $x_1(x_o)$. Reparallelizing the flow about any one of these singularities is somewhat like resetting the calendar when crossing the international dateline, except that a nonintegrable flow is generally not confined globally to a two dimensional analytic manifold.

We have reasoned that a "nonintegrable" flow is piecewise integrable: different sets of formulae of the form (4) hold *in principle* for consecutive finite time intervals $0 \leq t(x_o) < t(x_1)$, $t(x_1) \leq t < t(x_2)$, ... $t(x_{n-1}) \leq t < t(x_n)$, .... , giving geometric meaning to Poincaré's dictum [10] that a dynamical system is generally neither integrable nor nonintegrable but is more or less integrable. Nonintegrable flows are describable over arbitrarily-many consecutive time intervals by the simple formulae of the form (4) except at countably many singular points $x_1(x_o), x_2(x_o), ...$ , where the n-1 initial conditions $y_{io}$ and the integration constant D must be reset. The relevance for Takens's embedding theorem is discussed in [9].

**Deterministic chaos as simple dynamics**

We have often read over the last twenty years that deterministic chaos can explain complex phenomena, but without having had a definition of "complex". This was the point of view in the era when computers were used to try to study chaoic motions via numerical integrations without error control, based upon floating point forward integrations of chaotic dynamical equations (or by forward iterations of chaotic maps). We have since learned that uncontrolled numerical integrations can be avoided, and correspondingly that chaotic dynamics can be understood from a certain topologic point of view as relatively simple dynamics. This "new" approach (roughly ten years old) is the consequence of analytic studies of chaotic systems using controlled approximations via a purely digital method called "symbolic dynamics".

Symbol sequences are equivalent to digit strings in some base of arithmetic. Since we are going to talk about digit strings it is both wise and useful to begin with the idea of a computable number [20,21]. The reason for this is simple: "algorithmically random" numbers and sequences "exist" in the mathematical continuum but require infinite time and infinite precision for their definition, and therefore have no application to either experiment or computation.

By a computable number, we mean either a rational number or an algorithm that generates a digit expansion for an irrational number in some base of arithmetic, like the usual grade school algorithm for the square-root operation in base ten (the same algorithm also works in any other integer base). If we use computable numbers as control parameters and initial conditions, then the chaotic dynamical systems typically studied in physics and chemistry are computable, e.g. via (2) combined with analytic continuation. The Lorenz model (1b) provides one example. Systems of chemical kinetic equations provide other examples.

Seen from the perspective of computability, the local solution (2) of a dynamical system (1) that is digitized completely in some base of arithmetic defines an "artificial automaton", an abstract model of a computer. The digitized initial condition constitutes the program for the automaton. In a chaotic dynamical system the part of the program that directs the trajectory into the distant future is encoded as the end-string $\varepsilon_{N+1}...$ of digits in an initial condition $x_0 = .\varepsilon_1\varepsilon_2...\varepsilon_N...$ . For example, the binary tent map $x_n = f(x_{n-1})$,

$$f(x) = \begin{cases} 2x, & x < 1/2 \\ 2(1-x), & x > 1/2 \end{cases}, \quad (6)$$

can be rewritten and studied naturally in binary arithmetic by writing $x_n = .\varepsilon_1(n)\varepsilon_2(n)...\varepsilon_n(n)...$, with $\varepsilon_i(n) = 0$ or 1. The map (2) is then represented by the simple automaton [21]

$$\varepsilon_i(n) = \begin{cases} \varepsilon_{i+1}(n-1), & \varepsilon_1(n-1) = 0 \\ 1 - \varepsilon_{i+1}(n-1), & \varepsilon_1(n-1) = 1 \end{cases}. \quad (6b)$$

For every possible binary-encoded "computer program" $x_0 = .\varepsilon_1(0)\varepsilon_2(0)...\varepsilon_N(0)...$ this automaton performs only a trivial computation: either it reads a bit in the program, or else flips the bit and reads it, then moves one bit to the right and repeats the operation. The logistic map at the period doubling critical point [22], in contrast, is capable of performing simple arithmetic.

Unlike the binary tent map in binary arithmetic, most dynamical systems do not admit a "natural" base of arithmetic. The logistic map $f(x) = Dx(1-x)$ with D arbitrary and the Lorenz model are examples. The series solutions of these dynamical systems can still be rewritten as automata in any integer base of arithmetic, albeit in relatively cumbersome fashion. However, there is a systematic generalization of solution of the binary Bernoulli shift map $x_n = 2x_{n-1}$ mod 1 via binary arithmetic that sometimes works: symbolic dynamics. The symbolic dynamics of a chaotic dynamical system can be defined, and solved digitally at least in principle, if the map has a generating partition [23]. For the binary tent map (6) the generating partition, in generation n, consists of the $2^n$ intervals $l^{(n)} = 2^{-n}$ that are obtained by backward iteration of the entire unit interval by the map (a chaotic one dimensional map contracts intervals in backward iteration). Each interval in the generating partition can be labeled by an n-bit binary (L,R) address (L and R are defined in figure 3) called a symbol sequence, as is shown in figure 4. The symbol sequence tells us the itinerary of the map, for n forward iterations, for any initial condition that is covered by the interval $l^{(n)}(\varepsilon_1\varepsilon_2...\varepsilon_n)$ labeled by the n-bit address $\varepsilon_1\varepsilon_2...\varepsilon_n$, where $\varepsilon_i$ = L or R [21].

Excepting pathological cases where the contraction rate in backward iteration is too slow, an infinite length symbol sequence corresponds uniquely to an infinitely-precise initial condition. Given a symbol sequence, coarsegrained statistics for any number $N_n$ of bins in the generating partition ($N_n = 2^n$ for the binary tent map) can be obtained merely by reading the sequence while sliding an N-bit window one bit at a time to the right, as is indicated in figure 5. Clearly, orbital statistics depend on initial conditions, and it is very easy to construct algorithms for initial conditions whose orbital statistics do not mimic the uniform invariant density of the binary tent map (e. g., $x_o$ = .101001000100001... qualifies and follows from an obvious algorithm). I have explained elsewhere why "random" initial conditions may be a bad assumption for a dynamical system far from thermal equilibrium [9,21].

Because the binary tent map generates all possible infinite-length binary sequences (almost all of which are not computable via any possible algorithm [20]), we can use that map to generate *any* histogram that can be constructed in finitely-many steps, merely by a correct choice of initial conditions [21]. Many different initial conditions will allow the dynamical system to generate the same coarsegrained statistics because the precise ordering of L's and R's in a symbol sequence doesn't matter in determining the histograms.

Liapunov exponents depend strongly on initial conditions, a fact that is not brought out by concentration on excessively simple models like the symmetric tent map, or numerical attempts to extract "the largest Liapunov exponent" of a chaotic dynamical system like the Lorenz model. Chaotic dynamical systems like the Lorenz model or the logistic map generally generate an entire spectrum of Liapunov exponents (and therefore also a spectrum of largest Liapunov exponents). The easiest way to understand this is to solve for the generating partition and Liapunov exponents of the asymmetric tent map [21], where only simple algebra is needed.

We define a *class of initial conditions* to consist of all initial conditions that yield the *same* Liapunov exponent $\lambda$. Correspondingly, we can say that a class of symbol sequences defines a single Liapunov exponent. The Boltzmann entropy per iteration $s(\lambda)$ of all symbol sequences with the same Liapunov exponent $\lambda$ defines the fractal dimension $D(\lambda) = s(\lambda)/\lambda$ of that class of initial conditions [12,21], so that a chaotic dynamical system generally generates spectra of both Liapunov exponents *and* fractal dimensions.

Both critical [22] and chaotic [23] dynamical systems may generate a natural partitioning of phase space, the generating partition, but not *every* nonintegrable dynamical system defines a generating partition. If a deterministic dynamical system has a generating partition then the symbolic dynamics can in principle be solved and the long-time behavior can be understood qualitatively, without the need to compute specific trajectories algorithmically from the algorithmic construction of a specific computable initial condition. For example, one need only determine the possible symbol sequences and then read them with a sliding n-bit window in order to generate the statistics in the form of a hierarchy of histograms (figure 5). In other words, a high degree of "computational compressibility" holds even if the dynamical system is critical or chaotic.

Every chaotic dynamical system generates infinitely-many different classes of statistical distributions for infinitely-many different classes of initial conditions, and at most one of those distributions is differentiable (unlike the case of equilibrium statistical mechanics, there is no empirical evidence to suggest that nature far from equilibrium evolves from unknown initial conditions to generate differentiable distributions [9]). The generating partition, if it exists, uniquely forms the support of every possible statistical distribution and also characterizes the particular dynamical system (the intervals $l^{(n)} = 2^{-n}$ characterize the binary tent map and the binary Bernoulli shift). For a system with a generating partition, topologic universality classes can be defined that permit one to study the simplest system in the universality class [24]. The infinity of statistical distributions is topologically invariant and therefore can *not* be used to discern or characterize a particular dynamical system within a universality class [21].

For maps of the unit interval, both the symmetric and asymmetric logistic maps peaking at or above unity belong to the trivial universality class of the binary tent map [21] (where all possible binary sequences are allowed). The topologic universality class is described by figure 4, and is defined by the complete binary tree. Dynamical systems that generate complete ternary trees or incomplete binary trees, e.g., define other universality classes. The two dimensional Henon map belongs to the universality class of chaotic logistic maps of the unit interval peaking beneath unity. The simplest model in this topologic universality class is the symmetric tent map with slope magnitude between 1 and 2, and the class is defined by a certain incomplete binary tree [24].

In these systems the long-time behavior can be understood qualitatively and statistically *in advance*, so that the future holds no *surprises*: the generating partition and symbol sequences can be used to describe the motion at long times, to within any desired degree of precision $l^{(n)}$, and multifractal scaling laws (via the $D(\lambda)$ spectrum) show how finer-grained pictures of trajectories are related to coarser-grained ones. In other words, universality and scaling imply relatively simple dynamics in spite of the fact that the word "complex" has often been used to describe deterministic chaos.

**Complex dynamics**

Scale invariance based upon criticality has been suggested as an approach to "complex space-time phenomena" based upon the largely unfulfilled expectation of finding universal scaling laws, generated dynamically by many interacting degrees of freedom and yielding critical states independent of parameter-tuning [25,26], that are ubiquitous in nature. This is equivalent to expecting that nature is mathematically relatively simple.

From the standpoint of computable functions and computable numbers we can generally think of a deterministic dynamical system as a computer with the initial condition as the program [21]. Thinking of dynamics from this point of view, it has been discovered that there is a far greater and far more interesting degree of complicated behavior in nonlinear dynamics than either criticality or deterministic chaos: systems of billiard balls combined with mirrors [27,27b], and even two-dimensional maps [28], can exhibit universal computational capacity via formal equivalence to a Turing machine. A system of nine first order quasi-linear partial differential equations has been offered as a computationally-universal system [29]. A quasi-linear first order partial differential equation in n variables can be replaced by a linear one in n+1 variables. *Maximum computational complexity is apparently possible in systems of linear first order partial differential equations.* Such systems are nondiffusive but can describe damped-driven dynamics and wave propagation.

For a dynamical system with universal computational capability a classification into topologic universality classes is impossible [28]. Given an algorithm for the computation of an initial condition to as many digits as computer time allows, nothing can be said in advance about the future *either statistically or otherwise* except to compute the dynamics with controlled precision for that initial condition, iteration by iteration, to see what falls out: there is no computational compressibility that allows us to summarize the system's long-time behavior, either statistically or otherwise. In contrast with the case where topologic universality classes exist there is no tree-like organization of a hierarchy of periodic orbits, stable, marginally stable, or unstable, that allows us to understand the fine-grained behavior of an orbit from the coarse-grained behavior via scaling laws, or to look into the very distant future for arbitrary (so-called "random") initial conditions via symbolic dynamics. There can be no scaling laws that hold independently of a very careful choice of classes of initial conditions. We do not know whethe either fluid turbulence or Newton's three-body problem fall into this category.

Some degrees of complexity are defined precisely in computer science [30] but these definitions, based soley on computability theory, have not satisfied physicists [31,31b,32]. According to von Neumann [33] a system is complex when it is easier to build than to describe mathematically. Under this qualitative definition the Henon map is not complex but a living cell is. In earlier attempts to model biologic evolution [34,35] information was incorrectly identified as complexity. The stated idea was to find an algorithm that generates information, but this is too easy: the square root algorithm and the logistic map $f(x) = 4x(1-x)$ generate information at the rate of one bit per iteration from rational binary initial conditions.

There is no correct model of a dynamic theory of the evolution of biologic complexity, neither over short time intervals (cell to embryo to adult) nor over very long time intervals (inorganic matter to organic matter to metabolizing cells and beyond). There is no physico-chemical model of the time-development of different degrees of complexity in nonlinear dynamics. No one knows if universal computational capability is necessary for biologic evolution, although DNA molecules in solution apparently are able to compute [36], but not error-free like a Turing machine or other deterministic dynamical system.

Moore has speculated that computational universality should be possible in a certain kind of conservative three degree of freedom Newtonian potential flow [28], but so far no one has constructed an analytic example of the required potential energy. We do not yet know the minimum number of degrees of freedom necessary for universal computational capability in a driven-dissipative flow (a digital computer is a very high degree of freedom damped-driven dynamical system via electric circuit theory). Diffusive motion is time-irreversible ($U^{-1}(t)$ doesn't exist for diffusive motion), but arguments have been made that some diffusive dynamical systems may have an asymptotic limit that is reached asymptotically-fast, where the motion is non-diffusive and is even time reversible on a finite dimensional attractor [37,38,39], and is therefore generated on the attractor by a finite dimensional deterministic dynamical system (1). However, if a diffusive dynamical system (the Navier-Stokes equations, e.g.) can be shown to be computationally-universal then it will be impossible to discover a single attractor that would permit the derivation of scaling laws for eddy cascades in open flows, or in other flows, *independently of specific classes of boundary and initial conditions.*

With a computationally-universal (and therefore computable) dynamical system (1), given a specific computable initial condition $x_o$, both that initial condition and the dynamics can in principle be encoded as the digit string for another computable initial condition $y_o$. If the computable trajectory $y(t) = U(t)y_o$ could be digitally decoded, then we could learn the trajectory $x(t) = U(t)x_o$ for the first initial condition (self-replication without copying errors). *This maximum degree of computational complexity may be possible* *in low dimensional nonintegrable conservative Newtonian dynamics.* Some features of nonintegrable quantum systems with a chaotic classical limit (the helium atom, e.g.) have been studied using uncontrolled approximations based on the low order unstable periodic orbits of a chaotic dynamical system [40], but we have no hint what might be the behavior of a low dimensional quantum mechanical system with a computationally-complex Newtonian limit. Interacting DNA molecules obey the laws of quantum mechanics but the biologically-interesting case can not be reduced to a few degrees of freedom.

**Can new laws of nature emerge from studies of complicated motions? [42]**

The empirical discovery of mathematical laws of nature arose from the study of the simplest possible dynamical systems: classical mechanics via Galilean trajectories of apples and Keplerian orbits of two bodies (the sun and one planet) interacting via gravity, and quantum mechanics via the hydrogen atom. Is there any reason to expect that simplicity can be short-circuited in favor of complexity in the attempt to discover new mathematical laws of nature? Some researchers expect this to be possible, but without saying how [41]. Consider first an example from fluid dynamics where an attempt has been made to extract a simple law of motion from a complicated time series.

Fluid turbulence provides examples of complicated motions in both space and time in a Newtonian dynamical system of very high dimension. We know how to formulate fluid mechanical time evolution according to Newton's laws of motion, the Navier-Stokes equations, but infinitely many interacting degrees of freedom represented by second order coupled nonlinear partial differential equations are the stumbling block in our attempt to understand fluid turbulence mathematically. We do not understand coupled nonlinear partial differential equations of either the first or second order well enough to be able to derive any of the important features of fluid turbulence in either the finite or infinite Reynold's number limit from the Navier-Stokes equations in a systematic way that starts with the laws of energy and momentum transport and makes controlled, systematic approximations.

Can eddy-cascades in turbulent open flows [43] be understood by trying to build simpler mathematical models than the Navier-Stokes equations? So far, this goal remains nothing but an unfulfilled hope. Setting our sights much lower, is it possible to derive a mathematical law in the form of an iterated map that describes only the transition to turbulence, near criticality?

We have noted above that the binary tent map can generate all possible histograms that can be constructed simply by varying classes of initial conditions. *Statistics that are generated by an unknown dynamical system are therefore inadequate to infer the dynamical law that generates the observed statistical behavior* [21]. That is why, in any effort to derive a simplified dynamical system that describes either turbulence or the transition to turbulence, one cannot rely upon statistics alone. Instead, it is necessary to extract the generating partition of the dynamical system from the empirical data, *if there is a generating partition*.

Consider a low dimensional dynamical system that is described by an unknown iterated map defined finitely by a generating partition. With infinite precision and infinite time, it would be possible in principle to pin down the map's universality class and also the map, from a chaotic time series by the empirical construction of the generating partition. With finite precision and finite time one must always resort to some guesswork after a few steps in the hierarchy of unstable periodic orbits, which are arranged naturally onto a tree of some order and degree of incompleteness [24]. In practice one can discover at most only a small section of the tree and its degree of pruning, so if one is to narrow down the practical choices to a few topologic universality classes of maps the observational data must be extremely precise. Given the most accurate existing data on a fluid dynamical system near a critical point, the unique extraction of the universality class of an iterated map from a chaotic time series has yet to be accomplished without physically-significant ambiguity [44], demonstrating how difficult is the empirical problem that one faces in any attempt to extract an unknown law of motion from the analysis of complicated empirical data.

The method of topologic universality classes [21,23,24] is the only known way to study the long time behavior of a chaotic dynamical system systematically, meaning without the introduction of uncontrolled and uncontrollable errors. *For truly complex dynamical systems, therefore, our analysis suggests that the extraction of laws of motion from empirical data is a hopeless task.* This conclusion does not provide encouragement for experimental mathematicians who want to discover socio-economic or biologic laws of chaotic dynamics from raw statistics or the analysis of time series [45]. The alternative, to imagine that one could "guess" laws of nature without adequate empirical evidence or corresponding symmetry principles, would be to ignore the lessons of Archimedes and Galileo and revert to Platonism.

Einstein apparently became Platonic later in life, but Platonism was not Einstein's guiding light (or light-shade) during his generalization of Newton's theory of gravity, because that generalization is based upon a local invariance principle: no experiment can be performed to detect any difference between a linearly-accelerated frame of reference and the effect of a local gravitational field. This local symmetry principle was not accounted for by Newtonian theory, and motivated Einstein to discover a new set of gravitational field equations [46].

**Is socio-economic behavior (mathematically-)lawful?**

Is it reasonable, even in principle, to expect that mathematical laws of socio-economic or other mathematical laws of human behavior exist in any humanly-discernable form? Is it possible abstractly to *reduce* some aspects of human behavior to a set of universal formulae, or even to a finite set of more or less invariant rules? Many economists and system theorists [32,47,48], and even some sociologists [49,50], assume that this is possible.

By disregarding Galileo's historic and fruitful severing of the abstract study of inanimate motion from imprecise Aristotelian ideas of "motion" like youths alearning, acorns asprouting [2], and markets emerging, many mathematical economists have attempted to describe the irregularities of individual and collective human nature as if the price movements of a commodity, which are determined by human decisions and man-made political and economic rules, would define mathematical variables and abstract universal equations of motion analogous to ballistics and astronomy (deterministic models), or analogous to a drunken professor (stochastic models).

Mathematical economists often speak of the economy [48], which is determined by human behavior and man-made rules (and also in part by the weather, geology, and other limiting physical factors), as if the economy could be studied mathematically as an abstract dynamical system like the weather. In the latter case the equations of motion are known but cannot be solved approximately over large space-time regions by using floating point arithmetic on a computer without the introduction of uncontrollable errors. However, for specified boundary and initial conditions the weather is determined by the mathematical equivalent of many brainless interacting bodies that can not use intelligience to choose whether or not to obey the deterministic differential equations whose rigid mathematical rule they are condemned forever to follow. Chaos and complexity do not install either randomness, freedom of choice, or arbitrariness in the solutions of deterministic dynamical equations [21]. The absence of arbitrariness, or freedom of choice, is part of the key to understanding why mathematics works in physics but not in the socio-economic fields. Comparing the weather with socio-economic behavior is not a scientifically-sound theoretical analogy.

Contrary to certain expectations [51] and to recent extraordinary claims [52], there is no evidence to suggest that abstract dynamical systems theory can be used either to explain or understand socio-economic behavior. Billiard balls and gravitating bodies have no choice but to follow mathematical trajectories that are laid out deterministically, *beyond the possibility of human convention, invention, or intervention*, by Newton's laws of motion. The law of probability of a Brownian particle also evolves deterministically according to the diffusion equation *beyond the possibility of human convention, invention, or intervention.* In stark contrast, a brain that directs the movements of a body continually makes willful and arbitrary decisions at arbitrary times that cause it to deviate from and eventually contradict any mathematical trajectory (deterministic models) or evolving set of probabilities (stochastic models) assigned to it in advance. Given a hypothetical set of probabilities for a decision at one instant, there is no algorithm that tells us how to compute the probabilities correctly for later times, excepting at best the trivial case of curve-fitting at very short times, and then only if nothing changes significantly. Socio-economic statistics can not be known in advance of their occurrence because, to begin with, there are no known socio-economic laws of motion that are correct.

Economists stress that they study open systems, whereas physics concentrates on closed systems. This claim misses the point completely. We can describe and *understand* tornadoes and hurricanes mathematically because the equations of thermo-hydrodynamics apply, in spite of the fact that the earth's atmosphere is an *open* dynamical system. We can *not* understand the collapse of the Soviet Union or the financial crisis in Mexico on the basis of any known set of dynamics equations in spite of the fact that the world economy forms a *closed* financial system.

Mathematical-lawlessness reigns supreme in the socio-economic fields, where nothing of any social or economic significance is left even approximately invariant by socio-economic evolution, including the "value" of the Mark. This is the reason that artificial law ("law") must be used by governments and central banks in the attempt to regulate human behavior, both individually and collectively. Socio-economically, everything that is significant changes completely unregulated in the absence of police-enforced artificial law (the Roman method) or strong community traditions (the tribal method).

In the socio-economic fields there are no fundamental constants because *nothing is left invariant* by the time evolution. That nothing is left invariant is the same as saying that the system is not describable by mathematics: dynamical systems, even discrete ones [52b], have local conservation laws. Deterministic dynamical systems obey n-1 local conservation laws that prevent any external constraint from being imposed on the system. You can not "legislate" a change in the dynamics of a system that obeys a deterministic law of motion.

The division of observable phenomena into machine-like and not-machine-like behavior was made by Descartes [53]. In the Cartesian picture animals are supposed to behave more like machines, like robots that respond mechanically to stimuli. People, in contrast with robots, can reason and make decisions freely, or at least arbitrarily. Even the most illiterate or most stupid people can speak, can invent sentences creatively, and can behave unpredictably in other ways as well. The most intelligent dog, cat, or cow cannot invent intellectual complexity that is equivalent to a human language or a capitalist economy.

We should ask: why should *any* part of nature behave mathematically, simulating an automaton? Why does the mathematics of dynamical systems theory accurately describe the motions studied in physics, but not the "motions" (in Aristotle's sense) studied in economics, political science, psychology, and sociology? This question leads to Wigner's discussion of the "unreasonable effectiveness of mathematics" in describing the inanimate aspects of nature that physics traditionally studies, and that fields disconnected from physics have tried unsuccessfully to imitate merely by *postulating* laws of motion that do not pass the test of reproducibility of measurements.

It is necessary to realize that, in spite of Newton's scholastic style of presentation of his laws, which many mechanics text books unfortunately mimic, physics is neither postulatory nor axiomatic. Physics since Galileo is grounded in a deep interplay of empiricism and mathematical abstraction, and the reason that this mathematical interplay is at all possible is due to certain invariance principles (physics would be impossible in the absence of certain fundamental constants of nature; those constants reflect certain invariance principles).

**Reductionism, invariance principles, and laws of nature**

Reductionism is the arbitrary division of nature into laws of motion and initial conditions, plus "the environment". We must always be able to neglect "the environment" to zeroth order, because if *nothing* can be isolated then a law of motion can never be discovered. For example air resistance had to be negligible in order that Galileo could discover the law of inertia and the local law of gravity.

The empirical discovery of mathematical laws of motion that correctly describe nature is impossible in the absence of empirically-significant invariance principles, but there are no laws of nature that can tell us the initial conditions. Following Wigner, laws of motion themselves obey laws called invariance principles, while initial conditions are completely lawless [54]. *Why* must mathematical laws of motion that describe nature obey invariance principles?

*"It is not necessary to look deeper into the situation to realize that laws of nature could not exist without principles of invariance. This is explained in many texts of elementary physics even though only few of the readers of these texts have the maturity necessary to appreciate these explanations. If the correlations between events changed from day to day, and would be different for different points of space, it would be impossible to discover them. Thus the invariances of the laws of nature with respect to displacements in space and time are almost necessary prerequisites that it be possible to discover, or even catalogue, the correlations between events which are the laws of nature.*

E. P. Wigner in Symmetries and Reflections [54]

Nearly every elementary physics text shows that the experiments that Wigner had in mind are the parabolic trajectories of apples and blocks sliding down inclined planes, the two physical systems originally studied by Galileo in his empirical discovery of the local versions of Newton's first two laws of motion. Those discoveries would have been impossible in the absence of four fundamental invariance principles.

Without translational and rotational invariance in space and translational invariance in time (at least locally, on earth and within our solar system), simple *mathematical laws of motion* like the Keplerian planetary orbits and the Galilean trajectories of apples could not have been discovered in the first place. The experiments that are needed to discover the law of inertia are precisely *reproducible* because absolute position and absolute time are irrelevant as initial conditions, which is the same as saying that space is homogeneous and isotropic (space is locally Euclidean) and that the flow of time is uniform. The translational invariance of the law of inertia $d\mathbf{p}/dt=0$ means that the law of inertia can be verified regardless of where, in a tangent plane on earth, you perform the required experiment. The law of Galilean invariance is inherent in the law of inertia.

Socio-economic phenomena are not invariant in any empirically-discernable sense. Socio-economic time-development and the corresponding statistics depend upon absolute position and absolute time, which is the same (for all practical purposes) as admitting that socio-economic "motions" are not reducible to a well-defined dynamical system.

Dynamical laws of motion are *postulated* in economics, but the laws of physics are not mere postulates: mathematical laws of time-development come second, invariance principles come first. The law of inertia had to be discovered first (Galileo/Descartes) before Newton could write down his second order differential equation that generalizes Galileo's two local empirical laws, the law of inertia and the local law of gravity. Described from the standpoint of invariance and symmetry, the law of inertia is the foundation of all of physics: from it and Galileo's local law of gravity follow two of Newton's three laws of motion and his law of gravity as a generalization, when Kepler's first law and the action-reaction principle are used [12]. It is superficial and misleading to imagine that the law of inertia can be "derived" from Newton's second law merely by setting the net force equal to zero.

If absolute time and absolute position were relevant initial conditions then neither the law of inertia nor the local law of gravity would hold: identically prepared experiments would yield entirely different outcomes in different places and at different times. In this case there could have been no regularities discovered by Galileo, and no generalizations to universal laws of classical mechanics could have been proposed by Newton. Physics would, in that case, have remained Aristotelian and consequently would have evolved like economics, sociology, psychology, and political science: the study of a lot of special cases with no universal time-evolution laws that permit the prediction, or at least understanding, of phenomena over more than the short time intervals where curve-fitting sometimes "works", and with no qualitative understanding whatsoever of the phenomena underlying the observed "motions" and their corresponding statistics.

System theorists commonly assume that the economy operates like a dynamical system, the equivalent of an automaton that is too simple to simulate any kind of creative behavior, including the violation of politically-enforced laws as occurred during the collapse of the government of the former Soviet Union and the peasant rebellion in Chiapas. This is a strange assumption. Without human brains and human agreements based upon language, "laws" of economic behavior certainly could not exist. Dogs, cows and even peasants generally don't invent money-economies. In contrast, the available geological and astronomical evidence indicates that Newton's laws of motion held locally in our corner of the universe long before human languages emerged on earth.

Wigner considers that we can not rule out that "holistic" laws of nature (beyond general relativity, for example) might exist, but if so then we have no way to discover them. Reductionism can not explain *everything* mathematically, but reductionism is required in order to explain the phenomena that *can* be understood mathematically from the human perspective. Maybe an "oracle" would be required in order to discern the workings of a holistic law of motion.

Summarizing, universal laws that are determined by regularities of nature differ markedly from human-created systems of merely conventional behavior. The latter consist of learned, agreed-on, and communally- or politically-enforced behavior, which can always be violated by willful or at least clever people. People and groups who violate artificial law are sometimes called either "progressive" or "outlaw", depending on which social group does the labeling. In Wigner's language, all socio-economic initial conditions matter because of the lack of invariance, so that it is impossible to discover any underlying correlations that could be identified as mathematical laws of socio-economic "progress" (note that the idea of progress is also a "motion" only in the Aristotelian rather than in the Galilean sense).

**Darwinism and neo-Darwinism [55[1]]**

*"From a physicist's viewpoint, though, biology, history, and economics can be viewed as dynamical systems."*

P. Bak and M. Paczuski in <u>Complexity, Contingency, and Criticality</u> [52]

"Reductionism" (a better word is "science") is criticized by "holists" for not taking us far enough in our understanding of the world (see the introduction to ref. [32] and also [56]; see also any attempt by the so-called postmodernists to discuss science [57]). Some holists hope to be able to mathematize Darwinism in order to go beyond physics and chemistry (see discussions of "complex adaptable systems" [32]), but so far they have not been able use their invented dynamics models to predict or explain anything that occurs in nature. Physics and astronomy, since the divorce from Platonic mathematics and Aristotlelian "holism" in the seventeenth century, have a completely different history (or "evolution") than "political economy" and most of biology. "Emergence, selection, and adaptation" are buzz words used by Darwin-oriented holists (see ref. [58] for an alternative form of "holism"), while postmodernists like to toss around the notion of "a new paradigm for science". According to the postmodernists, "chaos" (which is merely a part of classical mechanics or chemical kinetics) is an example of "a new paradigm". "Paradigms" are very important for philosophers who have not understood science at the level of Galilean kinematics, and who can not distinguish science from pseudo-seience. Paradigms and "metaphors" are also important for people who know that a particular model doesn't represent what the researcher purports to study, but wants to claim that it does anyway.

---

[1] Kelly's book "Out of control" is a bible of "paradigms" of postmodernist "holistic" thought.

The Aristotelian dream of a holistic approach to physics, biology, economics, history, and other phenomena was revived by Bertanffly in 1968 [59] under the heading of system theory. System theory proposes to use mathematics to describe the time evolution of "the whole", like a living organism or a money-economy, but generally in the absence of adequate information about the local correlations of the connected links that determine the behavior of the whole.

I call attempts to quantify the Aristotelian style of thought "reductionist holism", or "holistic reductionism" because any mathematization whatsoever is an attempt at reductionism [42]. Quantification necessarily ignores all nonquantifiable qualities, and there are plenty of qualitative and quantitative considerations to ignore if we want to restrict our considerations to a definite mathematical model. Some physicists tend to believe that physics, which is successful reductionism (often with several-to-high decimal accuracy in agreement between theory and reproducible observations), provides the basis for understanding everything in nature, but only in principle [60].

There is no effective way to "reduce" the study of DNA to the study of quarks but this is not a failure of reductionism: both quarks and DNA are accounted for by quantum mechanics at vastly different length scales. In order to adhere to the illusion that reductionism might also be able account for biological and societal phenomena beyond DNA *in principle*, physicists must leave out of consideration everything that hasn't been accounted for by physics, which includes many practical problems that ordinary people face in everyday life. When sociologists [49,50] (who, unlike physicists, claim to interest themselves in the doings of ordinary people) try to follow suit but merely postulate or talk about dynamics "paradigms" in the absence of empirically-established invariance principles, then they *reduce* their considerations of society to groundless mathematical models, to artificial simulations of life that have nothing to do with any important quality of life.

Every computer simulation of a society or an economy is merely the creation of an abstract artificial and brainless society or an artificial and brainless economy. Mathematical simulations cannot adequately describe real societies and real economies although, through adequate politico-financial enforcement, which is truly a form of selection, we can be constrained to simulate some economist's simulation of society and economics. A money economy represents a selection based upon material resources and human needs, desires, and illusions. The idealized free market system described by Adam Smith's "invisible hand" represents a vague notion of autonomy, or self-regulation, inspired in part by Calvinism and in part by Watt's flywheel governor, but is in no scientific sense a "natural" selection.

Darwin's ideas of "natural selection, fitness, and adaptation" may appear to make sense in both sports and the socio-economic context of daily life but they are not scientifically-defined mathematical terms. That they remind us of the description of an organized market economy is not accidental: Darwin was strongly influenced during the cruise of the Beagle by his second reading of Malthus [61], who was both a protestant preacher and a worldly philosopher. Terms like "selection" and "adaptation" are reminiscent of Adam Smith's vague "invisible hand" rather than of scientifically well-defined processes like the dissociation and recombination of DNA molecules described by quantum mechanics or chemical kinetics.

In an attempt to model the origin of life, chemical kinetic equations have been used to try either to discover or to invent Darwinism at the molecular level [35] but the use of that terminology seems either superfluous or forced: a deterministic system of ordinary differential equations, whether chemical kinetic or not, can be described by the relatively precise, standard terminology of dynamical systems theory (stability, attractors, etc.). A stochastic system of chemical kinetic equations can be described by purely dynamic terminology combined with additional terms like "most probable distribution" and "fluctuations". There is far less reason to believe that Darwin's socio-economic terminology applies at the macromolecular level than there was, before 1925, to believe that the language of the Bohr model correctly described the motions of electrons relative to nuclei in hydrogen and helium atoms.

There are two main sources of Darwin's vague notion of "natural selection". The social-Darwinist origin of the phrase is Malthus's socio-economic doctrine, which derives from Calvinism [61] and can be traced through the late medieval revival of puritanism by Luther, Calvin, and Zwingli back to the neo-Platonist St. Augustine [6], who bequeathed to the west the notion of selection called "predestination". In "predestination" humans are divided completely arbitrarily into "the elect" and "the damned" (according to Luther, man is only an ass ridden by both God and the devil, with no choice whatsoever as to his ultimate fate [62]). Here, "selection" is *not* a mathematical idea that describes the time-evolution of a dynamical system. The second and *only* scientific motivation for Darwin's vague idea of "natural selection" came from plant and animal breeding, which he mislabeled as "artificial selection." Plant and animal breeding constitute the only *true* case of selection because they proceed via manipulating certain initial conditions in order to try to achieve a desired result.

Darwinists, true to their Aristotelian heritage, are condemned to argue endlessly to try to find out what their terminology means because that terminology is, from a scientific standpoint (empiric or theoretic), *completely undefined*.

The scientific foundation of organic evolution was established in Darwin's time by Mendel, who chose to become an Augustinian monk out of financial necessity [39] and was trained more in mathematics and physics than in biology. In contrast with Luther and Calvin, Mendel was not Augustinian in education and outlook: he was even a lecturer in experimental physics for a while, and approached the problem of heredity via isolation of cause and effect in the spirit of a physicist (or a good auto mechanic[2] ).

Darwin and his contemporaries, in contrast, accepted a holistic (or "integrated") picture of heredity that made the understanding of genetics impossible [64]. It was only after Mendel's reductionist discovery that some biologists began to dislodge themselves from the teleological notion of organic evolution as progress toward a goal predetermined by a selector (or read by an "oracle" capable of "infinite knowledge" of both future and past). By ignoring "the whole" in favor of the most important parts inferred from performing simple, controlled experiments, Mendel found the key that divorced the study of heredity from unsystematic tinkering and socio-economic doctrine and changed it into a precise mathematical science [64b]. Today, Darwinist concepts play a part in genetics research that is comparable to the role played by "waves" in high energy physics. "Wave-particle duality", rather than the Dirac-Feynman interpretation of quantum mechanics, is still taught in physics and chemistry courses, but you may scour the literature to no avail in an attempt to find reference to this cumbersome and unnecessary philosophic principle in particle physics research papers.

---

[2] Personally, I would not entrust my auto to a self-proclaimed "holist" for trouble-shooting prior to necessary repairs. See also Ginsburg [63b] for a nonmathematical alternative to holism in the social sciences.

Human history is narrative. This includes the statistics of socio-economic phenomena, which constitute only one very small part of the entire narrative, a quantitative part. There is no reason to expect that the uncontrolled approximations of system theory modelling can tell us as much, quantitatively or qualitatively, about social or individual behavioral phenomena as we can learn from experience and by reading history and novels (see [65] for an uncontrolled approximation to the description of some of the consequences of the unrestricted mechanization). The reason why it is illusory to expect to discover objective laws of human history, including the history ("time-evolution") of socio-economic development, was explained prosaically in 1952:

*"There can be no 'pure history'---history-in-itself, recorded from nobody's point of view, for nobody's sake. The most objective history conceivable is still a selection and an interpretation, necessarily governed by some special*

*interests and based on some particular beliefs. It can be more nearly objective if those interests and beliefs are explicit, out in the open, where they can be freely examined and criticized. Historians can more nearly approach the detachment of the physicist when they realize that the historical 'reality' is symbolic, not physical, and that they are giving as well as finding meanings. The important meanings of history are not simply there, lined up, waiting to be discovered."*

<div align="right">Herbert J. Muller, in <u>The Uses of the Past</u> [65]</div>

**One dimensional life**

*"The nineteenth century, in western Europe and North America, saw the beginning of a process, today being completed by corporate capitalism, by which every tradition which has previously mediated between man and nature was broken."*

<div align="right">John Berger, in <u>About Looking</u> [66]</div>

John Berger, in a very beautiful essay introducing the latest edition of <u>Pig Earth</u> [67], emphasizes what he calls the peasants' view of "circular time" in contrast with the abstract idea of linear time used in Newtonian mechanics. A related viewpoint was developed earlier by the Spengler [68], who was one is three historians who attempted to construct evidence for a grand scheme according to which human "history" evolves.

Following the anti-Newtonian Goethe, Spengler imagined human societies as "organisms" moving toward a "destiny". "Destiny" represents a vague idea of organic determinism that Goethe assumed to be in conflict with mechanistic time-evolution that proceeds via local cause and effect. "Destiny" was imagined to be impossible to describe via mathematical ideas, via Newtonian-style mechanism. In trying to make a distinction between global "destiny" and local cause and effect Spengler was not aware of the idea of attractors in dynamical systems theory, whereby time evolution mimics "destiny" but proceeds *purely mechanically according local cause and effect*. The Lie-Klein idea of invariance of geometry under coordinate transformations, the forerunner of Nöther's theorem on symmetry, invariance, and conservation laws in physics, may have inspired Spengler's attempt to compare entirely different cultures, widely separated in time and space, as they evolved toward "destinies" that he identified as fully-developed civilizations.

Spengler characterized western (European/North American) "civilization" in the following way: the entire countryside is dominated, Roman-style, by a few extremely overpopulated cities called megalopolises. Traditional cultures, derived from man's historic experience of wresting survival directly from nature, have been replaced by the abstract driving force of late civilization, the spirit of money-making. Spengler identified the transition from early Greek culture to late Roman civilization as an earlier example of the nearly "universal" evolution from local tribal culture to money-driven civilization.

In modern and postmodern civilization, in a single uncontrolled approximation, all traditions and ideas that interfere with "progress" defined as large-scale and efficient economic development are rejected as unrealistic or irrelevant in the face of a one-dimensional quantitative position whose units may be dollars or marks. The dialogue paraphrased below can be found on pg. 16 the book Complexity, Metaphors, Models, and Reality [32] about complex adaptable systems in biology, economics, and other fields. A, A', and A'', who are paraphrased, are theoretical physicists.

A: Why try to define measures of complexity? A measure of complexity is just a number and that doesn't tell you anything about the system. Assume that there's a particular state that you want to create, a slightly better state of the economy, for example. Suppose that you want to know how complicated that problem is to solve on a computer, and that you're able to characterize complexity. One of the proposals of A' for defining the complexity of a problem is 'what's the minimum amount of money you'd need in order to solve it?'

A'': The cost is proportional to computer time.

A: Then maybe the unit of complexity should be "money". If you're able to formalize the difficulty of solving the problem of making the economy slightly better, and you find out that you can measure its complexity in terms of dollars or yen, then that kind of measure would be extremely useful.

The prediction of a computable chaotic trajectory is limited, decimal by decimal or bit by bit, by computation time, but there are also integrable many body problems that are not complex but also require large amounts of marks or dollars. A'' also asserts in the same book Complexity (pg. 11) that low dimensional chaos is "not complex in a true sense: ... the number bits required for specification of where you are is highly limited." In part, this assertion is false: note that the binary specification of a single state $x_n$ in the logistic map $f(x) = 4x(1-x)$ requires precisely $N(n) = 2^n(N_o - 2) + 2$ bits, where $N_o$ is the number of bits in any simple initial condition $x_o = .\varepsilon_1...\varepsilon_{N_o}000...$ . If the string representing $x_n$ is arbitrarily truncated to $m \leq N(n)$ bits, then after on the order of m iterations the first bit (and all other bits) in $x_{n'}$, where $n' \approx n + m$, is *completely wrong* [7]. Multiplication of two finite binary strings of arbitrary length cannot be carried out on any fixed-state machine [5], and if multiplication is done incorrectly *at any stage* then after only a few more iterations the bits in $x_n$ cannot be known *even to one-bit accuracy*. I expect that the complexity of a dynamical system, like fractal dimensions and Liapunov exponents, can not be described by a single number.

*"Paradise...was..the invention of a relatively leisured class. ... Work is the condition for equality. ... bourgeois and Marxist ideals of equality presume a world of plenty, they demand equal rights before a cornucopia ... to be constructed by science and the advancement of knowledge. ... The peasant ideal of equality recognizes a world of scarcity ... mutual fraternal aid in struggling against this scarcity and a just sharing of what the world produces. Closely connected with the peasant's recognition, as a survivor, of scarcity is his recognition of man's relative ignorance. He may admire knowledge and the fruits of knowledge but he never supposes that the advance of knowledge reduces the extent of the unknown. ... Nothing in his experience encourages him to believe in final causes ... . The unknown can only be eliminated within the limits of a laboratory experiment. Those limits seem to him to be naive.*

<div style="text-align:right">John Berger, in <u>Pig Earth</u> [67]</div>

**Danksagung**

**References**

1. J. L. Casti (1990) <u>Searching for Certainty, What Scientists can know about the Future</u> (Wm. Morrow & Co., New York).

2. R. S. Westfall (1980) <u>Never at Rest, A Biography of Isaac Newton</u>, ch. 1 (Cambridge Univ. Pr., Cambridge).

3. P. A. Samuelson, (1976) <u>Economics</u> (McGraw-Hill, New York).

4. R. P. Feynman (1986) <u>Surely You're Joking Mr. Feynman</u>, (Bantom, New York) 308-17.

5. J. L. Casti (1992) <u>Reality rules: picturing the world in mathematics</u>, in two volumes (Wiley, New York).

6. Koestler, A. (1959) <u>The Sleepwalkers</u> (Macmillan, New York).

7. Hopkin, D., and Moss, B. (1976) <u>Automata</u> (Elsevier North-Holland).

8. M. Tabor et al (1991) in <u>What is Integrability?</u>, ed. by V. E. Zakharov (Springer-Verlag, Berlin).

9. McCauley, J. L. (subm. 5/1996) <u>Nonintegrability, chaos, and complexity.</u>

10. A. Wintner (1941) The Analytical Foundations of Celestial Mechanics (Princeton, Princeton) sect. 194-202 & 227-240.

11. V. I. Arnol'd, V. V. Kozlov, and A. I. Neishtadt (1993) Mathematical Aspects of Classical and Celestial Mechanics, in Dynamical Systems III, ed. by V.I. Arnold (Springer-Verlag, Heidelberg).

12. J.L. McCauley (forthcoming in 1996) Classical Mechanics: flows, transformations, integrability, and chaos (Cambridge Univ. Pr., Cambridge).

13. V.I. Arnol'd (1981) Ordinary Differential Equations (M.I.T. Press, Cambridge, Mass.).

14. E. Lorenz (1963) J. Atm. Sc. $\underline{20}$ 130.

15. H. Poincaré (1993) New Methods of Celestial Mechanics (AIP, Woodbury, NY).

16. L. P. Eisenhart (1961) Continuous Groups of Transformations (Dover, New York).

17. S.A. Burns and J.I. Palmore (1989) Physica $\underline{D37}$, 83.

18. P.J. Olver (1993) Applications of Lie Groups to Differential Equations (Springer-Verlag, New York) 30.

19. J. Palmore (1995) private conversation.

20. A. Turing (1937) Proc. Lon. Math. Soc. (2) $\underline{42}$, 230.

21. J.L. McCauley (1993) Chaos, Dynamics, and Fractals: *an algorithmic approach to deterministic chaos* (Cambridge Nonlinear Science Series 2, Cambridge).

22. M. J. Feigenbaum (1988) J. Stat. Phys. $\underline{52}$, 527.

23. P. Cvitanovic, G. Gunaratne, and I. Procaccia 1503 Phys. Rev. A$\underline{38}$ (1988).

24. G. H. Gunaratne (1990) in _Universality beyond the onset of chaos_, in _Chaos: Soviet and American Perspectives on Nonlinear Science_, ed. D. Campbell (AIP, New York).

25. G. Grinstein (1995) in _Scale Invariance, Interfaces, and Non-Equilibrium Dynamics_, ed. A. Mckane et al (Plenum, New York).

26. J. Krug (1994) in _Modern Quantum Field Theory II_, ed. S. R. Das et al (World, Singapore).

27. E. Fredkin and T. Toffoli (1982) Int. J. Theor. Phys. **21** 219.

27b. C. Bennett (1982) Int. J. Theor. Phys. **21** 905.

28. C. Moore (1990) Phys. Rev. Lett. **64** 2354.

29. S. Omohundro (1984) Physica **10D** 128.

30. M. R. Garey and D.S. Johnson (1979) _Computers and Intractability: A Guide to the Theory of NP-Completeness_ (Freeman, San Francisco).

31. C. Bennett (1986) Found. Phys. **16** 585.

31b. S. Lloyd and H. Pagels (1988) Annals of Phys. **188**, 186.

32. G. A. Cowan, D. Pines, and D. Meltzer, editiors (1994) _Complexity, Metaphors, Models, and Reality_ (Addison-Wesley, Reading).

33. J. von Neumann (1966) _Theory of Self-Reproducing Automata_ (Univ. of Ill., Urbana) pp. 47, 48.

34. M. Eigen und R. Winkler-Oswatitsch (1987) _Stufen zum Leben,_ (R. Piper, München).

35. B. -O. Küppers (1985) _The Molecular theory of Evolution_ (Springer, Heidelberg).

36. R.J. Lipton (1995) Science **268** 542.

37. P. Constantin, C. Foias, O. P. Manley, and R. Temam (1985) J. Fluid Mech. **150** 427.

38. C. Foias, G. R. Sell, and R. Temam (1988) J. Diff. Eqns. **73** 309.

39. P. Constantin, C. Foias, B. Nicolaenko, and R. Teman (1980) <u>Integral and Inertial Manifolds for dissipative Partial Differential Equations</u> (Springer-Verlag, New York).

40. H. Friedrich (Apr. 1992) Physics World 32-36.

41. J. L. Casti (1995) Complexity <u>1</u>, 12.

42. J. L. McCauley (1996) <u>Simulations, complexity, and laws of nature</u> (refereed twice and accepted by editor John Casti for the Santa Fe Institute journal "Complexity", on 2/9/96 and then much later rejected on 5/4/96 because of "hostile" and "unprintable" reactions by four "commentators").

43. J. L. McCauley (1991) in <u>Spontaneous Formation of Space-Time Structures and Criticality</u>, ed. D. Sherrington and T. Riste (Kluwer, Dordrecht).

44. A. Chhabra, R. V. Jensen, and K. R. Sreenivasan (1989) Phys. Rev. A<u>40</u>, 4593.

45. E. E. Peters (1991) <u>Chaos and Order in the Capital Markets</u> (Wiley, New York).

46. A. Einstein (1953) <u>The Meaning of Relativity</u>, Princeton Univ. Pr., Princeton.

47. J. L. Casti and A. Karlqvist (1991) <u>Beyond Belief: Randomness, Prediction, and Explanation in Science</u> (CRC Press, Boca Raton).

48. P. W. Anderson, K. J. Arrow, and D. Pines, eds. (1988) <u>The Economy as an Evolving Complex System</u> (Addison-Wesley, Redwood City).

49. R. L. Ackoff (1974) <u>Redisigning the Future</u> (Wiley-Interscience, New York).

50. R. Collins (1988) <u>Theoretical Sociology</u> (Harcourt, Brace, Janovich, San Diego).

51. D. Ruelle (July, 1994) Physics Today, pp. 24-30.

52. P. Bak and M. Paczuski (1995) Complexity, Contingency, and Criticality, subm. to Proc. Nat. Acad. Sci..

52b. J. Palmore (1995) Chaos, Solitons, and Fractals 5 1397.

53. N. Chomsky (1968) Language and Mind, Harcourt Brace Janovich (New York).

54. E. P. Wigner (1967) Symmetries and Reflections (Univ. Indiana Pr., Bloomington).

55. K. Kelly (1994) Out of Control (Addison-Wesley, Reading) ch. 22.

56. H. Morowitz (1995) Complexity 1 4.

57. J.- F. Lyotard (1993) The Postmodern Condition: A Report on Knowledge, trans. from fr. by G. Bennington and B. Massumi (Univ. of Minn., Minneapolis).

58. P. Bak (1994) Self-Organized Criticality: A Holistic View of Nature in ref.

59. L. von Bertalanffy (1968) General Systems Theory (G. Braziller, New York).

60. S. Weinberg (10/5/1995) Reductionism Redux, New York Review of Books, pp. 39-42.

61. R. M. Young (1985) Darwin's Metaphor (Cambridge Univ. Pr., Cambridge).

62. H. A. Oberman (1982) Mensch zwischen Gott und Teufel (Severin und Seidler Ver., Berlin).

63. R. Olby (1985) Origins of Mendelism, 2nd Ed. (Univ. of Chicago Pr., Chicago).

63b. C. Ginsburg (1989) Clues, Myths, and the Historical Method transl. from Italian (Johns Hopkins Pr., Baltimore) pp. 96-125.

64. P. J. Bowler (1989) The Mendelian Revolution (Johns Hopkins, Baltimore).

64b. The Human Genome Project (1992) Los Alamos Science 20.

65. K. Vonnegut (1952) Player Piano (Avon Books, New York).

66. H. J. Muller (1952) <u>The Uses of the Past</u> (Oxford Univ. Pr., New York).

66. J. Berger (1991) <u>About Looking</u> (Vintage Pr., New York).

67. J. Berger (1992) <u>Pig Earth</u> (Vintage Pr., New York).

68. O. Spengler (1922) <u>Untergang des Abendlandes</u> (C. H. Bech'sche Verlagsbuchhandlung, München).

**Figure Captions**

1. Samuelson's question: Is it a bird or an antelope? Answer: neither, it's a continuous line between two points plus a closed curve that, unlike both birds and antelopes, is topologically equivalent to a straight line plus a circle (from Samuelson [3]).

2. Successive maxima $z_n$ of a numerically-computed time series $x_3(t)$ for the Lorenz model are plotted against each other (from McCauley [21]). The drawing of a single continuous curve through all of these points would violate the time-reversibility.

3. Assignment of the symbols L and R for a unimodal map (from McCauley [21]).

4. The complete binary tree defines the topologic universality class of the binary tent map, and all unimodal maps of the unit interval that peak at or above unity and contract intervals exponentially fast in backward iteration (from McCauley [21]).

5. An n-bit sliding window (shown for n = 1, 2, and 3) is used to read a section of a binary symbol sequence in order to discover the corresponding orbital statistics, described by histograms with $2^n$ bins (from McCauley [21]).